\documentclass[prl,aps,twocolumn,showpacs,superscriptaddress]{revtex4-1}
\usepackage{graphics}
\usepackage{epsfig}
\usepackage{amsmath}
\usepackage{amsfonts}
\usepackage{amssymb}
\usepackage{graphicx}
\usepackage{color}
\usepackage{tabularx}
\usepackage{txfonts}
\usepackage{multirow}

\begin{document}
\title{Charge-state dependent spin-orbit coupling and quantum phase transitions in Ir-Ru oxides}
\author{Kuldeep Kargeti}
\affiliation{Department of Physics, Bennett University, Greater Noida-201310, Uttar Pradesh, India}
\author{Bidyut Mallick}
\affiliation{Department of Applied Sciences, Galgotias College of Engineering and Technology, Greater Noida, India}
\author{Vladislav Borisov}
\affiliation{Department of Physics and Astronomy, Uppsala University, Box 516, SE-751 20 Uppsala, Sweden}
\author{Sk. Soyeb Ali}
\affiliation{Department of Physics, Bennett University, Greater Noida-201310, Uttar Pradesh, India}
\author{Johan Hellsvik}
\affiliation{PDC Center for High Performance Computing, KTH Royal Institute of Technology, SE-100 44 Stockholm, Sweden}
\author{Olle Eriksson}
\affiliation{Department of Physics and Astronomy, Uppsala University, Box 516, SE-751 20 Uppsala, Sweden}
\affiliation{Wallenberg Initiative Materials Science for Sustainability, Uppsala University, SE-751 21 Uppsala, Sweden}
\author{S. K. Panda}
\email{Electronic address: swarup.panda@bennett.edu.in}
\affiliation{Department of Physics, Bennett University, Greater Noida-201310, Uttar Pradesh, India}

\begin{abstract}
The competition between kinematic, relativistic, and Coulombic interactions has spurred intense experimental and theoretical investigations in iridium-based oxides. Its electronic structure is mostly understood in terms of the spin-orbital coupled effective $J$-state ($J_{eff}$). However, the role of the Ir charge state in shaping the strength of effective spin-orbit coupling and defining the stability of the $J_{eff}$ ground state has not been thoroughly explored. We argue here that the Iridium-Ruthenium triple perovskites, Ba$_3$MRuIrO$_9$ (M = Li, Mg and In), are of particular interest in this regard. Using ab-initio theory, we show here that the nominal charge states of Ir can be tuned from +6 (5$d^3$) to +4 (5$d^5$) by choosing non-magnetic 'M' ions as Li (+1), Mg(+2) and In (+3), while the Ru ions always remain in nominal +5 (4$d^3$) charge-state. This variation modulates the influence of the spin-orbit coupling (SOC) which is found to be negligible in Ba$_3$LiRuIrO$_9$, moderate in Ba$_3$MgRuIrO$_9$ and determining in Ba$_3$InRuIrO$_9$. Our analysis classifies Ba$_3$LiRuIrO$_9$ as a band-insulator, Ba$_3$MgRuIrO$_9$ as a SOC and correlation driven insulator that does not conform to the commonly expected $J=0$ ground state and Ba$_3$InRuIrO$_9$ as $J_{\rm eff} = 1/2$ Mott-Hubbard insulator. In the data reported here, the correlational electronic structure theory results in sizeable magnetic moments of both Ru and Ir atoms in these systems and atomistic spin-dynamics simulations capture the experimental N\'eel temperature for Ba$_3$LiRuIrO$_9$ and Ba$_3$MgRuIrO$_9$ and provide evidence for a phase transition for Ba$_3$InRuIrO$_9$ when T $\to$ 0 K, to a multi-valley magnetic state with strong magnetic frustration. The theory identifies that strong SOC in Ba$_3$InRuIrO$_9$ induces bond-dependent magnetic couplings with significant Dzyaloshinskii-Moriya (DM) interaction, strong symmetric anisotropic exchange, and finite in-plane single-ion anisotropy. The realization of such strong anisotropic interactions helps to stabilize a particularly complex energy landscape of Ba$_3$InRuIrO$_9$, that opens up for exotic magnetic quantum phases like quantum spin liquid. Thus, by comparing the electronic structure and magnetism of iso-structural iridates with different Ir-charge states, we provided a theoretical framework to demonstrate the structural and electronic conditions that drive the deviations from conventional magnetic ordering, facilitating the emergence of exotic quantum phases.
\end{abstract}

\maketitle

\section{INTRODUCTION}
In the last decade, iridium based oxides (iridates) brought substantial attention due to their potential of offering a completely new venue for correlated physics and emergent quantum phenomena that arise from a cooperative effect of large crystal field, electron-electron interaction (Hubbard $U$) and strong spin-orbit coupling (SOC)~\cite{Balents_Review,Sr3MIrO6,Li2IrO3,katukuri2015strong,ko2015charge,Na2irO3,MIT_Dimension,QSL_Na4Ir3O8,QSL_Ba3IrTi2O9,TI_Na2IrO3_1,TI_Na2IrO3_2,Pesin2010}. 
In most of the iridates, Ir ions are octahedrally coordinated by O-ions, leading to a well separated $t_{2g}$ states which behave like $p$-orbitals ($l_{\rm eff}$ = 1). Due to heavy weight of Ir-ions and the partially filled $d$-orbitals, strong SOC as well as significant Hubbard $U$ are realized~\cite{Pesin2010,JHalf_Kim,sr2IrO4-sc}. The strong SOC is responsible for the splitting of  $t_{2g}$ bands ($l_{\rm eff}$ = 1) into J$_{\rm eff}$ = 3/2 and J$_{\rm eff}$ = 1/2 states~\cite{JHalf_Kim} in
single particle picture by a gap of $\frac{3\lambda}{2}$, $\lambda$ being the SOC strength. In such a scenario, the electronic structure of iridates where Ir exhibit nominal 4+ charge state ($5d^5$) could be understood in terms of fully filled J$_{\rm eff}$ = 3/2 quartets and a half filled J$_{\rm eff}$ = 1/2 states that further open up a Mott-like gap due to the presence of Hubbard $U$ as was first reported in Sr$_2$IrO$_4$~\cite{JHalf_Kim}. This microscopic model in the presence of strong SOC also naturally provide a spin-orbital J$_{\rm eff}$ = 3/2 state for Ir with 6+ charged state ($5d^3$), and $J$ = 0 nonmagnetic band-insulating state (no magnetic moments or magnetic response) when Ir exhibits 5+ charge state ($5d^4$). However, deviation from the $J$ = 0 nonmagnetic state and resultant complicated magnetic behavior have been extensively reported in real materials~\cite{Bhowal_2021,terzic2017evidence,meetei2015novel,khaliullin2013excitonic}. Hence, the $J = 0$ non-magnetic ground state in  iridates with $d^4$ electron configuration remains a controversial topic and is still under intense debate. For instance, double-perovskite iridates Sr$_2$YIrO$_6$~\cite{PhysRevLett.112.056402} and  Ba$_2$YIrO$_6$~\cite{PhysRevB.96.064436} hosts magnetic moments and a long-range magnetic ordering despite Ir being in a pentavalent state ($5d^4$). The reason of such magnetic ground state in these systems are attributed to the non-cubic (distorted octahedra) crystal field as well as intermediate SOC strength which can not stabilize the $J$ = 0 ground state. In this context, an interesting issue that did not receive much attention is how the effective SOC strength, $\lambda \vec{L}\cdot\vec{S}$ ($\vec{L}$ is the orbital moment and $\vec{S}$ is the spin moment) changes with the charge state of Ir ions in Iridium based oxides. In fact, the magnetism and electronic properties of iridates having Ir$^{6+}$ ($5d^3$) have received much less attention compared to pentavalent and tetravalent iridates~\cite{Ir6+PhysRevB.100.064418,Ir6+yang2015crystal,Ir6+kayser2013crystal,Ir6+DEMAZEAU1993479}. 
\begin{figure}[t]
    \begin{center}
     \includegraphics[width=0.99\columnwidth]{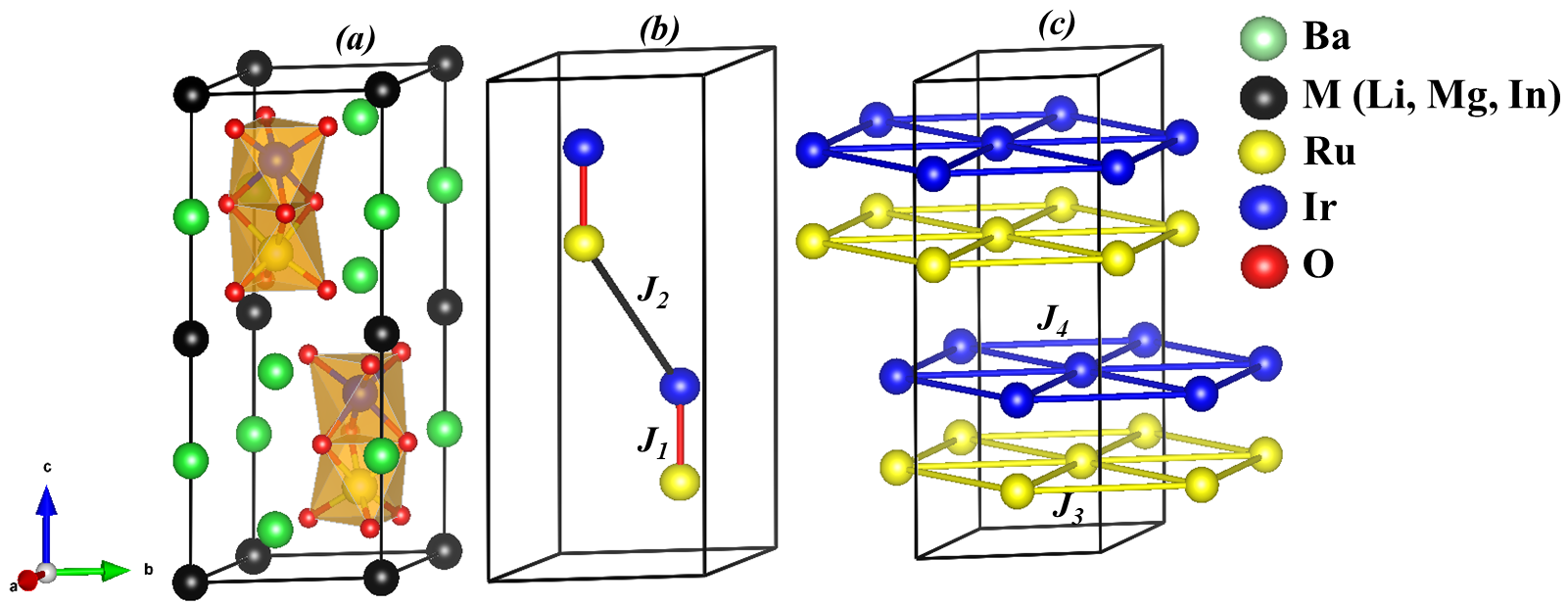}
    \caption {(a) The unit cell of the triple perovskite Ba$_3$MRuIrO$_9$ (where M can be Li, Mg, or In). The crystal structure features Ru-Ir dimers, where Ru and Ir ions occupy face-sharing octahedra. (b) A schematic representation of the magnetic exchange interactions between nearest-neighbor (NN) and next-nearest-neighbor Ru and Ir ions. These interactions play a crucial role in determining the material's magnetic properties. (c) Illustration of the Ru-Ru and Ir-Ir magnetic interactions within the hexagonal network formed by these ions in the a-b plane. This hexagonal arrangement introduces the potential for magnetic frustration in the system.}
    \label{crystal}    
    \end{center}
\end{figure}
\par 
The above discussion indicates that a thorough investigation is required to understand the critical role of the iridium charge state in stabilizing the spin-orbit-coupled $J_{\rm eff}$ states in iridates. The mixed ruthenium-iridium triple-perovskite series~\cite{lufaso2005crystal}, represented by the chemical formula Ba$_3$MRuIrO$_9$ (M = Li, Mg, In), offers a unique platform to explore this intriguing phenomenon. In this series of compounds, the Ir charge state could be tuned by suitably choosing the non-magnetic ions "M".  The crystal structure of these materials are displayed in Fig. \ref{crystal}(a). They possess a hexagonal crystal structure with space-group P6$_3$/mmc. The "M" ions are located at the corner-sharing octahedral site and the face-sharing octahedral sites are occupied by a disordered mixture of Ru and Ir-ions. In order to understand the preferred positions of Ru and Ir ions, we have computed the energies of various possible configurations within density functional theory formalism and found that Ir and Ru preferred to form a face-sharing structural dimer as displayed in Fig. \ref{crystal}(b). Such arrangements of Ir and Ru agree with the recent observation of dimer formation in similar compounds~\cite{kumar2024electronic}. Another interesting aspect of these materials are that both Ru and Ir ions form a hexagonal network in the $a-b$ plane (see Fig. \ref{crystal}(c)) which potentially could introduce magnetic frustration provided the magnetic interactions between them are antiferromagnetic in nature. The magnetic susceptibility data~\cite{lufaso2005crystal} of Ba$_3$MRuIrO$_9$ shows different behaviour as a function of "M". A signature of antiferromagnetic ordering is observed for Ba$_3$LiRuIrO$_9$ and Ba$_3$MgRuIrO$_9$ at 180 K and 100 K, respectively, while Ba$_3$InRuIrO$_9$ does not posses any long-range ordering down to the lowest temperature measured. The ground state electronic structure of these systems are reported to be insulating in nature~\cite{lufaso2005crystal}. 
\par 
In order to understand the diverse magnetic properties of this family of compounds and most importantly to investigate the influence of the Ir charge state on modulating the strength of effective spin-orbit coupling, we carried out detailed theoretical investigations based on density functional theory + Hubbard U (DFT+U) calculations with and without SOC. Our results from DFT+U demonstrate that Ru always remain pentavalent (4d$^3$), while
nominal charge state of Ir in these three compounds comes
out to be 6+ (Ba$_3$LiRuIrO$_9$), 5+ (Ba$_3$MgRuIrO$_9$) and 4+
(Ba$_3$InRuIrO$_9$), respectively. We find that despite having large $\lambda$ due to its large atomic number $Z$ ($\lambda \sim Z^4$ in the atomic limit), the influence of SOC ($\lambda \vec{L}\cdot\vec{S}$) is negligible for Ru-ions (4d$^3$) in all the three compounds since the Ru-$t_{2g}$-states are exactly half-filled, resulting in a quenching of the orbital moments. For the same reason, the effective SOC for Ir$^{6+}$ ($t_{2g}^3$) in Ba$_3$LiRuIrO$_9$ is found negligible and the insulating state arises purely due to the local moment formation (Hund's coupling).  However, the orbital moment and the effective SOC is significant in Ba$_3$MgRuIrO$_9$ where Ir has nominal $t_{2g}^4$ occupation and the insulating state is realized only in the presence of SOC. The strength of effective SOC is largest in Ba$_3$InRuIrO$_9$ where Ir exhibit 4+ charge state ($t_{2g}^5$) and it is a good candidate for relativistic $J_{\rm eff}$ = $\frac{1}{2}$ Mott-Hubbard insulator. Thus, we show that effective SOC gradually enhances as we go from nominal $t_{2g}^3$ to $t_{2g}^5$ state in Ir and a a subtle interplay between crystal field, Hund's coupling, electronic correlation and spin-orbit interaction underpins the electronic and magnetic properties of mixed Ru-Ir triple perovskites. Further, we outlined a theoretical framework to demonstrate the necessary structural and electronic condition thar drive the deviation from conventional magnetic ordering, facilitating the emergence of exotic quantum phases, i.e. quantum spin liquid (QSL). 
\begin{figure}[t]
    \centering
    \includegraphics[width=0.99\columnwidth]{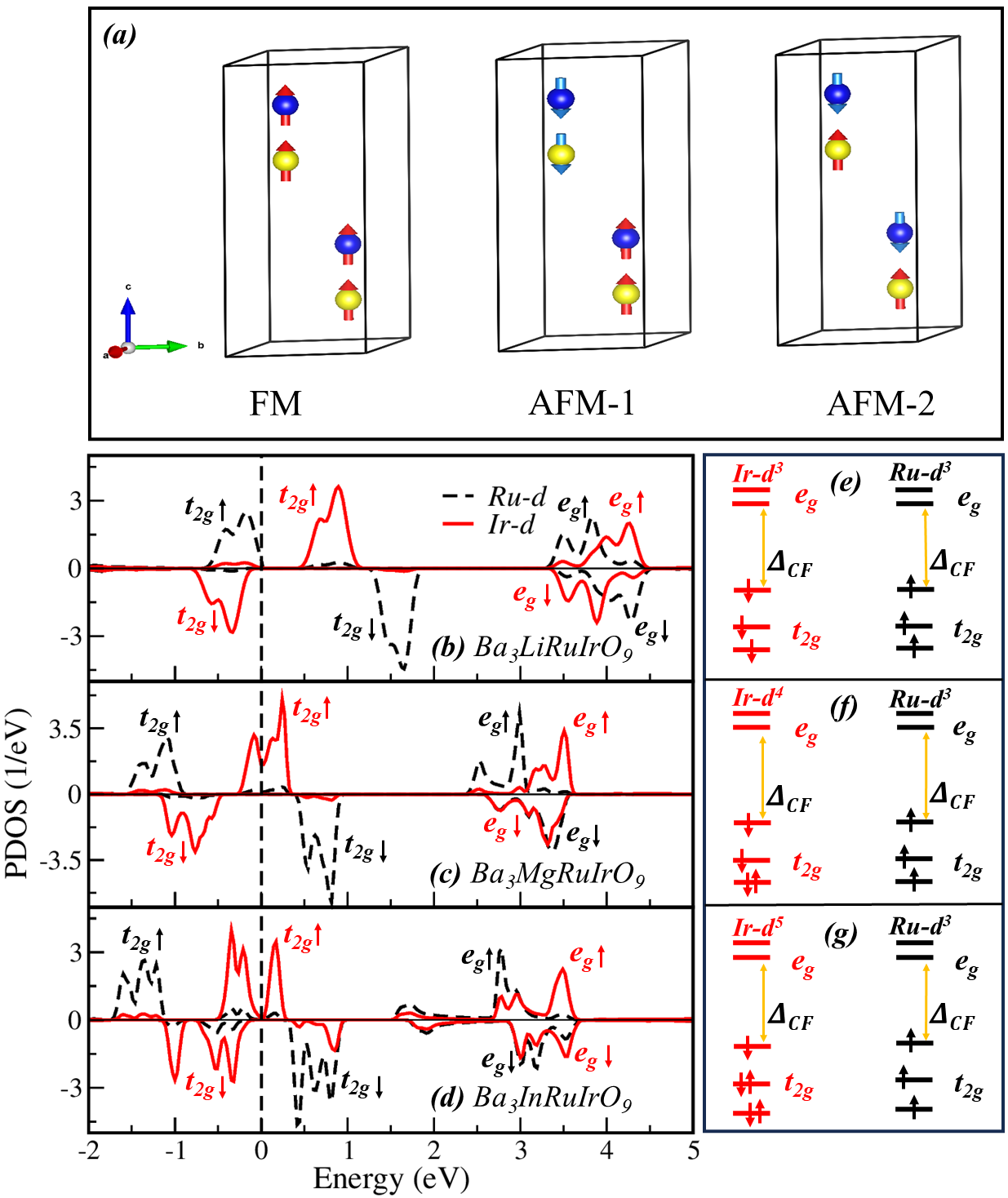}
    \caption {(a) The schematic of various possible magnetic configurations within the unit-cell to determine the lowest energy magnetic state. Panels (b), (c), and (d) present the Ru-4d and Ir-5d partial density of states (PDOS) for the Li, Mg, and In substituted compounds, respectively. Corresponding schematics in panels (e), (f), and (g) illustrate the crystal field splitting and the nominal electron occupations of the Ru and Ir d-states as obtained in PDOS. These schematics aid in visualizing the relationship between charge state, orbital filling, and the resulting magnetic behavior.}
    \label{GGA+U-PDOS}
\end{figure}

\par 
\begin{table*}
\centering
\caption{Energies corresponding to three different configurations obtained from GGA+U+SOC calculations with computed spin ($\mu_{S}$) and orbital ($\mu_{L}$) moments for Ru and Ir ions in the lowest energy state. Ratio of orbital to spin moment is also listed. The energies are written with respect to the ferromagnetic (FM) state.} 
\label{moment}
\begin{tabular}{|c|ccc|cc|cc|}
\hline
\multirow{2}{*}{} & \multicolumn{3}{c|}{Energy (meV/u.c.)}                                  & \multicolumn{2}{c|}{Moments on Ru in $\mu_B$}            & \multicolumn{2}{c|}{Moments on Ir in $\mu_B$}            \\ \cline{2-8} 
 & \multicolumn{1}{c|}{FM} & \multicolumn{1}{c|}{AFM1} & AFM2 & \multicolumn{1}{c|}{$\mu_{S}(\mu_{L})$} & $\mu_{L}/\mu_{S}$ & \multicolumn{1}{c|}{$\mu_{S}(\mu_{L})$} & $\mu_{L}/\mu_{S}$ \\ \hline
$Ba_{3}LiRuIrO_{9}$       & \multicolumn{1}{c|}{0.00} & \multicolumn{1}{c|}{-320.04} & -482.36 & \multicolumn{1}{c|}{1.77 (0.01)} & 0.01  & \multicolumn{1}{c|}{1.24 (0.01)} & 0.01  \\ \hline
$Ba_{3}MgRuIrO_{9}$       & \multicolumn{1}{c|}{0.00} & \multicolumn{1}{c|}{-95.19}  & -222.78 & \multicolumn{1}{c|}{1.84 (0.05)} & 0.03  & \multicolumn{1}{c|}{0.71 (0.31)} & 0.44  \\ \hline
$Ba_{3}InRuIrO_{9}$       & \multicolumn{1}{c|}{0.00} & \multicolumn{1}{c|}{-11.87}  & 0.00    & \multicolumn{1}{c|}{1.84 (0.02)} & 0.01  & \multicolumn{1}{c|}{0.42 (0.45)} & 1.10  \\ \hline
\end{tabular}
\end{table*}
\par 
\section{METHODS}
The electronic structure calculations presented in this study were conducted within the framework of density functional theory (DFT) using three distinct approaches. First, calculations were performed with a plane-wave basis as implemented in the Vienna Ab initio Simulation Package (VASP)~\cite{vasp1,vasp2}, employing projector augmented wave (PAW) potentials~\cite{PAW}. Second, the full-potential linearized augmented plane wave (FP-LAPW) method was utilized via the WIEN2K code~\cite{wien2k}. Lastly, the full-potential linear muffin-tin orbital (FP-LMTO) method, implemented in the RSPT code \cite{Codefplmto,FPLMTO}, was also employed. These complementary methods provided cross-validation of the computational results, thereby reinforcing the reliability of the conclusions. Exchange and correlation effects were treated using the GGA+U (generalized gradient approximation + Hubbard $U$) approach, where the Hubbard $U$ term was included to account for strong electronic correlations \cite{GGA_PBE,GGA+U}. The $U$ values for the Ru and Ir $d$-states were set to 3 eV and 2.0 eV, respectively, with a Hund's exchange parameter of 0.8 eV for both ions, consistent with values reported in prior studies~\cite{ba3niir2o9,Ru-U}. The Brillouin zone (BZ) integration was performed using the tetrahedron method with a $10 \times 10 \times 4$ k-mesh. To determine the lowest-energy arrangement of Ru and Ir atoms, ionic relaxation was carried out within the unit cell, minimizing forces to within 10$^{-3}$ eV/$\AA$. The effects of spin-orbit coupling (SOC) on the valence states were incorporated using a second-variational method with scalar relativistic wave functions. 
\par
\noindent\textbf{Inter-site magnetic exchange:}
The isotropic symmetric exchange interactions $J_{ij}$ are determined using the full-potential linearized muffin-tin orbital (FPLMTO) basis in the RSPt code \cite{Codefplmto,FPLMTO} through the well-known Lichtenstein-Katsnelson-Antropov-Gubanov (LKAG) methodology, which utilizes the Green’s function formalism. In this context, the concept is to connect the interaction between two spins with the change in energy that results from a minor perturbation in the magnetic state. In practice, this mapping is based on the magnetic force theorem, which indicates that the total energy variation of the electronic subsystem can be described solely in terms of changes in occupied single-particle energies~\cite{magnetic-force1,magnetic-force2}. The generalized formula for the intersite exchange parameters is presented in the equation below:
\begin{equation}
J_{i j}=\frac{\mathrm{T}}{4} \sum_n \operatorname{Tr}\left[\hat{\Delta}_i\left(i \omega_n\right) \hat{G}_{i j}^{\uparrow}\left(i \omega_n\right) \hat{\Delta}_j\left(i \omega_n\right) \hat{G}_{j i}^{\downarrow}\left(i \omega_n\right)\right]
\end{equation}
The quantities expressed in the above expression are the onsite spin splitting $\Delta_i$, the spin-dependent intersite Green’s function $G_{ij}$, and $\omega_n$ are the $n^{th}$ Fermionic Matsubara frequency respectively. 
To gain insights into more anisotropic magnetic interactions such as Dzyaloshinskii-Moriya (DM) interactions and symmetric anisotropic exchange terms in the spin Hamiltonian, it is necessary to extend this method to the relativistic framework, which has already been accomplished using the Korringa-Kohn-Rostoker (KKR) Green-function technique. A comprehensive explanation of the implementation of the Korringa–Kohn–Rostoker (KKR) method in RSPt is available in Refs.~\cite{RSPt,RevModPhys.95.035004}. The application of this methodology to various transition metal systems, as demonstrated in our previous work~\cite{LaCrS3} and other reported studies~\cite{Borisov}, has consistently validated the accuracy of the computed magnetic interaction parameters. This reliability supports the adoption of this approach in the present study.
\par
\noindent\textbf{Spin dynamics simulations:}
To comprehend the characteristics of magnetic ordering in the system, a spin Hamiltonian, as outlined in eq.~\ref{spin-hamil}, was developed for the three systems being examined. Utilizing the determined micromagnetic parameters presented in the spin Hamiltonian, we conducted micromagnetic simulations at finite temperature, as implemented in the UPPASD code~\cite{UppASD1,UppASD2}. The temperature was varied from 0 to 300 K, aligning with the range of parameters studied experimentally. Micromagnetic simulations were performed on a $20 \times 20 \times 20$ supercell with periodic boundary conditions for each system by solving the Landau-Lifshitz-Gilbert (LLG) equation~\cite{spindynamics}. The magnetization dynamics outlined by the LLG equations can be represented as,
\begin{equation}
    \frac{\partial \mathbf{m}_i}{\partial t}=-\frac{\gamma}{1+\alpha^2}\left(\mathbf{m}_i \times \mathbf{B}_i+\frac{\alpha}{m} \mathbf{m}_i \times\left(\mathbf{m}_i \times \mathbf{B}_i\right)\right),
\end{equation}
In this equation, $\gamma$ represents the gyromagnetic ratio, while $B_i$ denotes the effective magnetic field that each atom $i$ experiences, determined by the partial derivative of the Hamiltonian H (eq.~\ref{spin-hamil}) concerning the local magnetic moment $m_i$. The scalar phenomenological Gilbert damping parameter $\alpha$ is set to 1.0, facilitating the system's progression toward thermal equilibrium. A typical time step for solving the differential equations is $\Delta(t)$ = 0.1 femtoseconds, or 10$^{-16}$ seconds, which allows the achievement of the equilibrium state. Micromagnetic simulations are initiated from a randomly assigned magnetic configuration at a temperature of 300 K and then gradually cooled to 0 K, simulating the process of annealing. 
\par 
\section{RESULTS \& DISCUSSION}
\noindent\textbf{Electronic structure from GGA+U:}
It is widely believed that all Ir based oxides possess large SOC \cite{SOC-ming2018spin,SOC-ye2018covalency,SOC-panda2015electronic,SOC-pesin2010mott}. However, this work is aiming to understand the impact of effective SOC on the electronic and magnetic properties of iridates for various possible charge states of Ir-ions. To find out the charge states of Ir in this triple-perovskite family of compounds, we first analyse their electronic structure obtained from the GGA+U calculations without incorporating SOC. We considered three different possible combinations of intra and inter-dimer couplings within the unit-cell as displayed in Fig.~\ref{GGA+U-PDOS} (a) and computed their total energies. These results suggest that the AFM-2 magnetic structure, where both intra- and inter-dimer configurations are antiferromagnetic, is the lowest in energy for all three systems. In this magnetic state, we analysed the orbital and site-projected DOS (see Fig.~\ref{GGA+U-PDOS} (b)-(d)) of Ir-$d$ and Ru-$d$ orbitals to find out their nominal occupations and the relevant orbitals responsible for the magnetism in these materials. The PDOS reveals that $e_g$ orbitals of both Ru and Ir appear much above the Fermi energy and they are well separated from the $t_{2g}$ states, demonstrating a large crystal field splitting. This is common for 4$d$ and 5$d$ elements due to their extended nature which is responsible for strong bonding with the ligand states. For all the compounds, the Ru-$t_{2g}$-majority states are completely filled up, while minority $t_{2g}$ states appear above Fermi energy. Thus we observe a robust $S$=$\frac{3}{2}$ ($d^3$ nominal occupations) spin-state of Ru for all the mixed Iridium-Ruthenium triple perovskites. This is also evident from the computed spin-moment of Ru which almost remain unaltered as we change the materials. The magnitude of spin-moments on each Ru atom are 1.77 $\mu_B$ in Ba$_3$LiRuIrO$_9$, 1.84 $\mu_B$ in Ba$_3$MgRuIrO$_9$ and 1.84 $\mu_B$ in Ba$_3$InRuIrO$_9$. The deviation from the ionic value (3.0 $\mu_B$) occurs due to the delocalised nature of $4d$-states. In contrast, a significant change in spin-moment on Ir-ion is noticed upon changing the composition. It is 1.24  $\mu_B$ in Ba$_3$LiRuIrO$_9$, 0.71 $\mu_B$ in Ba$_3$MgRuIrO$_9$ and 0.42 $\mu_B$ in Ba$_3$InRuIrO$_9$. Such change in Ir-spin moments could be understood in terms of their nominal occupation. Since we have chosen three triple perovskite where the non-magnetic "$M$"-ions have nominal charge state of 1+ ($M$=Li), 2+ ($M$=Mg), 3+ ($M$=In) and Ru exhibit a robust nominal 5+ charge state ($d^3$), it is thus obvious that Ir-ions will change the nominal charge state from 6+ ($d^3$) to 4+ ($d^5$) to maintain the charge balance in each compound. The decrement in spin-moments on increasing nominal occupations is observed as Ir ions exhibit low-spin state where all the electrons remain in the $t_{2g}$ manifold and $e_g$ states appear far above Fermi energy with large $t_{2g}$-$e_g$ crystal field. This is evident from the PDOS of Ir-$d$ states as seen in Fig.~\ref{GGA+U-PDOS}(b)-(d). The Ir-$e_g$ states of both spin-channels are seen in the energy range of 3 to 4 eV above Fermi energy. For all the compositions, the minority Ir-$t_{2g}$ states are completely filled, while the majority $t_{2g}$ states are empty in Ba$_3$LiRuIrO$_9$ (Fig.~\ref{GGA+U-PDOS}(b)) and are partially occupied in Ba$_3$MgRuIrO$_9$ (Fig.~\ref{GGA+U-PDOS}(c)) and Ba$_3$InRuIrO$_9$ (Fig.~\ref{GGA+U-PDOS}(d)). The obtained scenario in terms of the occupation of Ru and Ir-$d$ orbitals is schematically explained in Fig.~\ref{GGA+U-PDOS}(e),(f) and (g). It demonstrates that the low-energy physics of this class of materials is governed by the Ir-$t_{2g}$ states whose occupations change from $t_{2g}^3$ to $t_{2g}^5$, providing an unique opportunity to study the charge state dependency of SOC phenomena in $l_{\rm eff} = 1$ state within the same class of materials. We also note that our GGA+U calculations are able to find an insulating ground state for Ba$_3$LiRuIrO$_9$ whose origin could be attributed to the large octahedral crystal-field, local moment formation and the occupation of $d$-orbitals. However, GGA+U wrongly predict metallic ground state for Ba$_3$MgRuIrO$_9$ and Ba$_3$InRuIrO$_9$. Thus, it is essential to assess the role of SOC on the electronic and magnetic properties of these three materials. \\
\par
\begin{figure*}[t]
    \includegraphics[width=1.99\columnwidth]{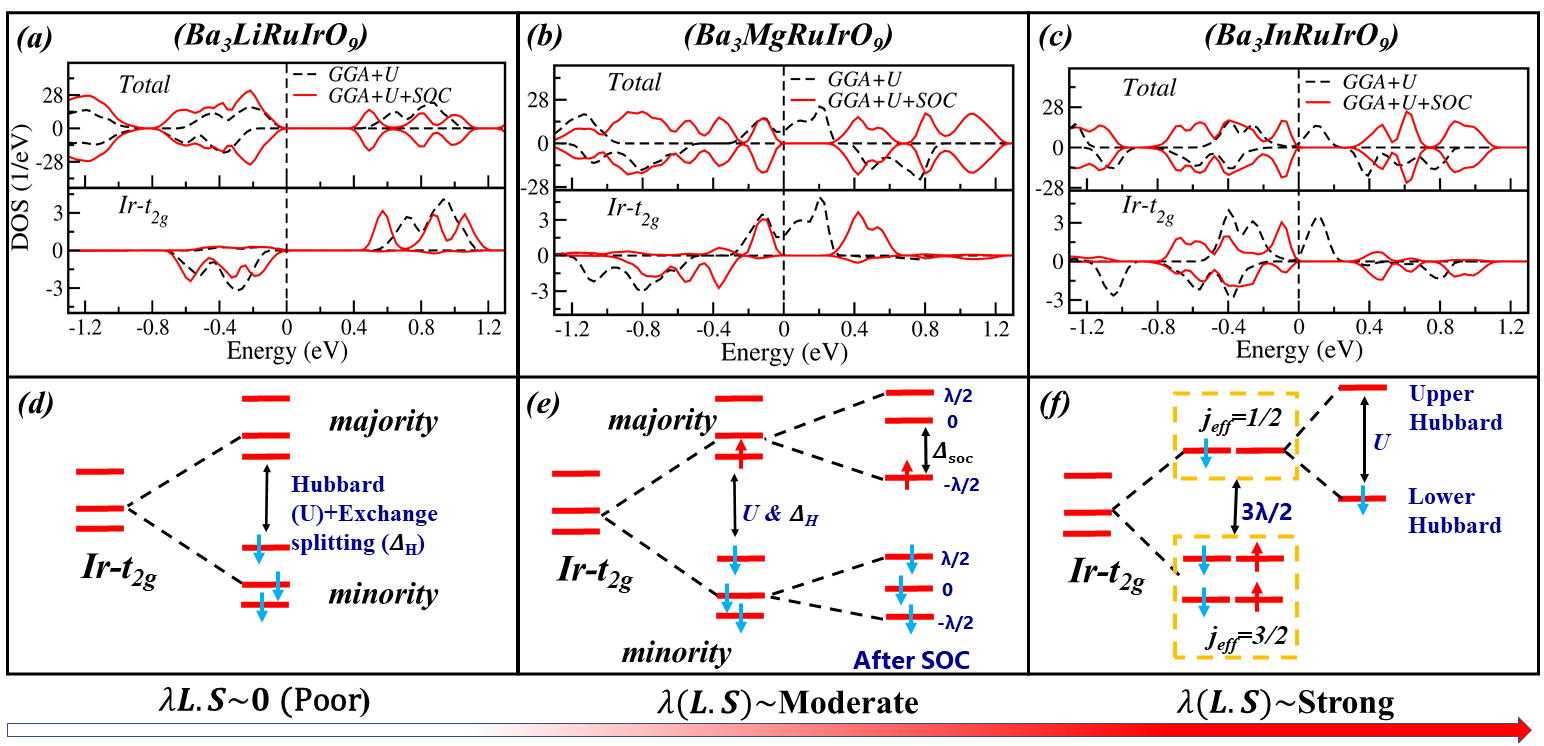}
    \caption {Comparison of total and Ir-$t_{2g}$ partial density of states (PDOS) obtained from GGA+U and GGA+U+SOC calculations for (a) Ba$_3$LiRuIrO$_9$, (b) Ba$_3$MgRuIrO$_9$, and (c) Ba$_3$InRuIrO$_9$. The upper panels depict how the inclusion of spin-orbit coupling (SOC) influences the electronic structure, particularly the redistribution of spectral weight and the formation of energy gaps. The lower panels provide schematics illustrating the insulating behavior in each material, corresponding to (d) negligible SOC in Ba$_3$LiRuIrO$_9$, where the insulating state arises due to local moment formation and crystal-field effects, (e) moderate SOC in Ba$_3$MgRuIrO$_9$, leading to a SOC-driven insulating state, and (f) strong SOC in Ba$_3$InRuIrO$9$, resulting in a relativistic $J_{eff} = 1/2$ Mott-Hubbard insulator. These results demonstrate the critical role of SOC strength in modulating the electronic structure of these materials.}
    \label{GGA+U+SOC-PDOS}
\end{figure*}
\noindent\textbf{Influence of spin-orbit coupling on electronic structure and magnetism:} In view of the above and to understand the role of Ir-charge state (and nominal occupations of $t_{2g}$ orbitals) in tuning the strength of effective SOC, we carried out GGA+U simulations including SOC (GGA+U+SOC approach). From the computed total energies (see Table~\ref{moment}), we see that the lowest energy magnetic state within the limit of our consideration remains AFM-2 for Ba$_3$LiRuIrO$_9$ and Ba$_3$MgRuIrO$_9$. However, for Ba$_3$InRuIrO$_9$, we observe that AFM-1 becomes the lowest energy state upon inclusion of SOC. We note here that in the absence of SOC, our GGA+U calculation shows AFM-2 as lowest energy state for all the considered compounds. Hence it is evident that SOC has strongest influence on the nature of magnetic coupling in Ba$_3$InRuIrO$_9$ where Ir posses nominal 4+ ($5d^5$) charge state. These observations could be further understood in terms of the magnitude of orbital moments, particularly from the ratio of the orbital-to-spin magnetic moment ($\frac{\mu_{L}}{\mu_{S}}$) which is considered to be a crucial quantity to understand the strength of effective spin-orbit interaction. 
As displayed in Table~\ref{moment}, the orbital moments of Ru ions are found to be negligibly small in all three compounds. This is expected as Ru possesses half-filled $t_{2g}$ orbitals (nominal $t_{2g}^3$ occupations) where orbital moments are nearly completely quenched, giving rise to a negligible contribution of Ru-ions to the effective SOC in this family of compounds. For the Ir ions in Ba$_3$LiRuIrO$_9$, orbital moment (0.01 $\mu_B$) is also nearly fully quenched which could be attributed to the half-filled $t_{2g}$ states alike Ru. Thus, we find that owing to the large octahedral crystal-field and filling of $d$-orbitals, the resultant SOC effect is nearly zero in Ba$_3$LiRuIrO$_9$. 
The computed orbital moment of Ir is much larger in Ba$_3$MgRuIrO$_9$ (0.31 $\mu_B$) and Ba$_3$InRuIrO$_9$ (0.42 $\mu_B$).  
Hence, the $\frac{\mu_{L}}{\mu_{S}}$ for Ir comes out to be 1.10 in Ba$_3$InRuIrO$_9$ which is two times larger compared to Ba$_3$MgRuIrO$_9$. 
We note here that $\frac{\mu_L}{\mu_S}$ is directly related to the 
Land\'e $g$-factor~\cite{PhysRev.76.743} 
which is often experimentally measured using techniques like x-ray magnetic circular dichroism (XMCD), electron paramagnetic resonance (EPR) to determine the strength of SOC. 
Thus, we find that among these three iso-structural iridates with similar chemical formula, the effective SOC is strongest in Ba$_3$InRuIrO$_9$, moderate in Ba$_3$MgRuIrO$_9$ and nearly zero in Ba$_3$LiRuIrO$_9$. 
\begin{table}[t]
\caption{Energies of different magnetization axis with respect
to the energy of $c$- axis magnetization for all the three systems. These calculations are performed within GGA+U+SOC approach with different spin quantization axes.}
\label{MAE}
\begin{tabular}{|l|l|l|l|}
\hline
Spin axis & $Ba_{3}LiRuIrO_{9}$       & $Ba_{3}MgRuIrO_{9}$       & $Ba_{3}InRuIrO_{9}$       \\ \hline
          & Energy (meV/f.u.) & Energy (meV/f.u.) & Energy (meV/f.u.) \\ \hline
$c$-axis       & 0.00              & 0.00              & 0.00              \\ \hline
$b$-axis       & 3.7              & 15.0              & -18.9             \\ \hline
$a$-axis       & 3.7              & 15.0              & -18.9             \\ \hline
\end{tabular}
\end{table}
\par
In order to provide further credence to this understanding, we have also estimated magnetocrystalline-anisotropy (MAE) which is defined as the difference between the total energies due to change in spin-quantization axis. It is generally understood that the primary origin of the large MAE in bulk materials is strong SOC ($\lambda \vec{L}\cdot\vec{S}$) which tries to align the moments with the magnetic field and thus the total energy depends on the orientation of the moments in the lattice. We note here that if the SOC constant ($\lambda$) is too small or orbital moment ($\mu_L$) is quenched, the coupling between the spins ($\vec{S}$) and their orbital motion ($\vec{L}$) is less pronounced (poor effective SOC), and in such cases, moments can rotate freely with respect to the field, resulting in weak MAE. Our results as summarized in Table~\ref{MAE} show that the MAE is highest for Ba$_3$InRuIrO$_9$ and lowest in Ba$_3$LiRuIrO$_9$. In fact, the easy axis of magnetization is $c$ for both Ba$_3$LiRuIrO$_9$ and Ba$_3$MgRuIrO$_9$ and it changes to easy-plane ($a$-$b$) in Ba$_3$InRuIrO$_9$. These results clearly suggest that charge state dependent nature of effective SOC in Ir based oxides. This contradicts the general belief of considering all Ir based compounds as high SOC materials and explaining their properties based on $J_{\rm eff}$ picture which originate due to strong coupling between spin and orbital momentum. Thus, we observe that charge state of Ir put a stability limits of spin-orbit-coupled $J_{\rm eff}$ states in iridium based oxides.

\begin{figure}
    \includegraphics[width=0.99\columnwidth]{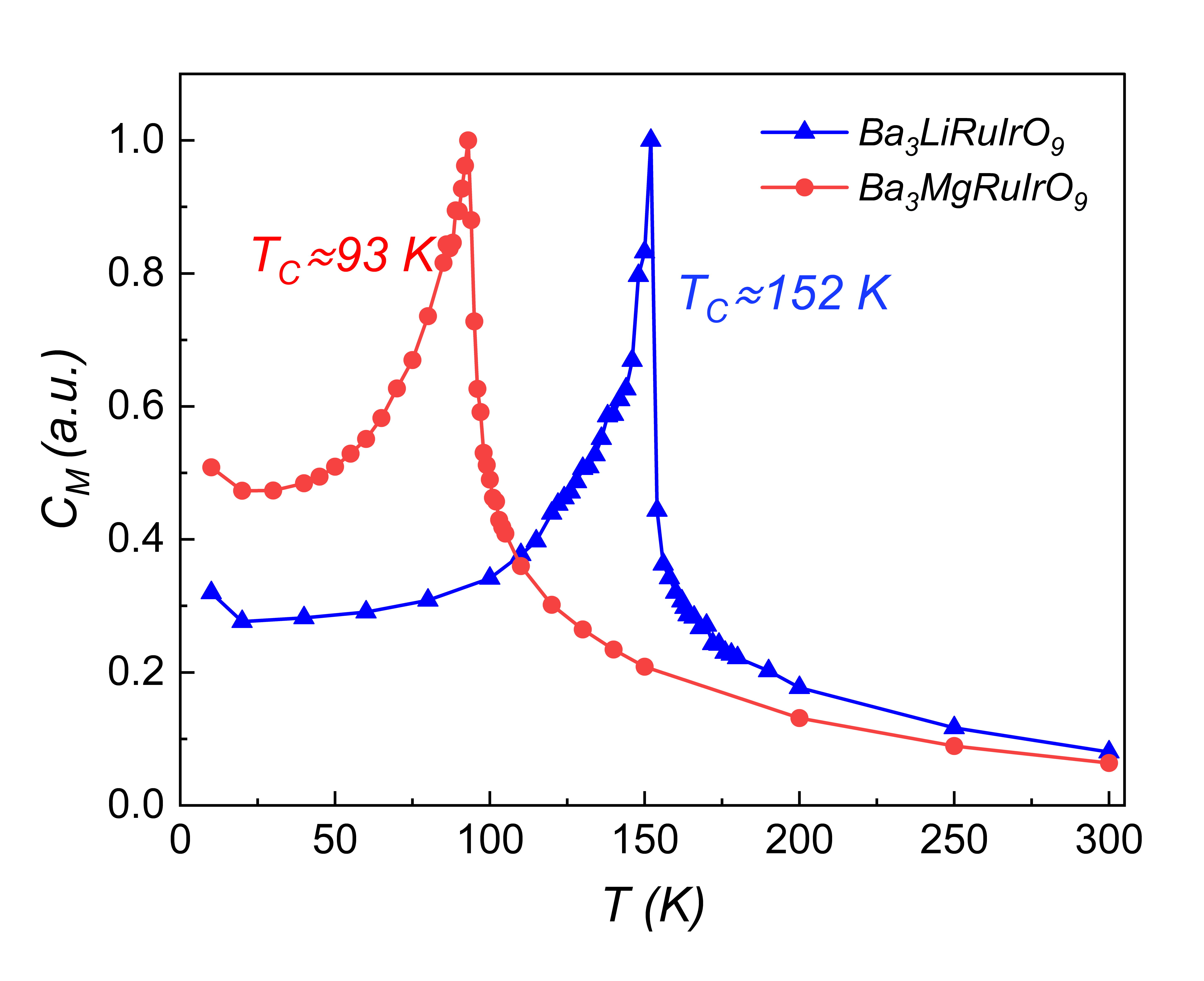}
    \caption{Magnetic contribution to the heat capacities of Ba$_3$LiRuIrO$_9$ and Ba$_3$MgRuIrO$_9$ computed using atomistic spin dynamics simulations. The presence of sharp peaks in the heat capacity data provides clear evidence for the emergence of long-range magnetic order in these compounds. The calculated transition temperatures (Tc) are found to be 152 K for Ba$_3$LiRuIrO$_9$ and 93 K for Ba$_3$MgRuIrO$_9$, in good agreement with experimental observations.}
   \label{ASD_results_1}
\end{figure}

\begin{table*}
\centering
\caption{Inter-site exchange coupling strengths derived from GGA+U+SOC Calculations. Classification of coupling types as antiferromagnetic ($J < 0$) or ferromagnetic ($J > 0$) with implications on magnetic ordering. Exchange couplings $J$s are illustrated in Fig.~\ref{crystal}.}
\begin{tabular}{|c|c|c|cc|cc|cc|}
\hline
\multirow{2}{*}{Interactions} & \multirow{2}{*}{\begin{tabular}[c]{@{}c@{}}No. of Neighbors\\ ($z_i$)\end{tabular}} & \multirow{2}{*}{Bonds} & \multicolumn{2}{c|}{$Ba_3LiRuIrO_9$}         & \multicolumn{2}{c|}{$Ba_3MgRuIrO_9$}         & \multicolumn{2}{c|}{$Ba_3InRuIrO_9$}        \\ \cline{4-9} 
                              &                                                                                  &                        & \multicolumn{1}{c|}{$J_{ij}$ (meV)}    & $\mid z_iJ_{ij}/J_1 \mid$ & \multicolumn{1}{c|}{$J_{ij}$ (meV)}    & $\mid z_iJ_{ij}/J_1 \mid$ & \multicolumn{1}{c|}{$J_{ij}$ (meV)}   & $\mid z_iJ_{ij}/J_1 \mid$ \\ \hline
$J_1$                            & 1                                                                                & Ir-Ru                  & \multicolumn{1}{c|}{-12.64} & 1          & \multicolumn{1}{c|}{-9.60} & 1          & \multicolumn{1}{c|}{16.91}  & 1          \\ \hline
$J_2$                            & 3                                                                                & Ir-Ru                  & \multicolumn{1}{c|}{-7.88}  & 1.87       & \multicolumn{1}{c|}{-3.26}  & 1.02       & \multicolumn{1}{c|}{-0.16} & 0.03       \\ \hline
$J_3$                            & 6                                                                                & Ru-Ru                  & \multicolumn{1}{c|}{-0.86}  & 0.41       & \multicolumn{1}{c|}{-0.97}  & 0.61       & \multicolumn{1}{c|}{-1.37} & 0.49       \\ \hline
$J_4$                            & 6                                                                                & Ir-Ir                  & \multicolumn{1}{c|}{-1.17}  & 0.56       & \multicolumn{1}{c|}{-0.65}   & 0.41       & \multicolumn{1}{c|}{-0.10} & 0.03       \\ \hline
\end{tabular}
\label{Isotropic}
\end{table*}
\par 
In order to further asses the effect of SOC on electronic structure, we have compared the total and orbital projected DOS for Ir-$t_{2g}$ orbitals, computed within GGA+U and GGA+U+SOC approaches  (Fig.~\ref{GGA+U+SOC-PDOS}). Since our GGA+U results (Fig.~\ref{GGA+U-PDOS}) suggest that the low energy physics is dominated by the $t_{2g}$ states and $e_g$ states appear far above Fermi energy, we have only focused on the spectral function around Fermi energy. As expected, SOC has a very small impact on the electronic states in Ba$_3$LiRuIrO$_9$ (see Fig.~\ref{GGA+U+SOC-PDOS} (a)). This results further provide credence to our conclusion that effective SOC is less dominating in Ba$_3$LiRuIrO$_9$ where Ir is in nominal $d^3$ (+6 charge state) state. The crystal-field splitting, local exchange field combined ($\Delta_H$) with correlation effects (Hubbard $U$) are sufficient to account for the insulating ground state in this system as schematically explained in Fig. \ref{GGA+U+SOC-PDOS}(d)). However, the scenario is completely different for Ba$_3$MgRuIrO$_9$ and Ba$_3$InRuIrO$_9$ where insulating state is obtained only after the inclusion of SOC as shown in the comparative total density of states plot of Fig.~\ref{GGA+U+SOC-PDOS}(b) and (c). The PDOS of Ir-$t_{2g}$ orbitals in Ba$_3$MgRuIrO$_9$ (lower panel of Fig.~\ref{GGA+U+SOC-PDOS}(b)) reveals that Ir-$t_{2g}$ minority channel is fully occupied and lies well below the Fermi level while majority channel spectral weight is reduced at the Fermi level resulting in a gap opening. The magnitude of the gap comes out to be 0.33 eV. Thus, from half-metallic behaviour as seen in GGA+U calculations to a pure insulating state due to presence of SOC in GGA+U+SOC approach is observed in Ba$_3$MgRuIrO$_9$. The computed electronic structure as well as experimental observation of magnetic transition are not consistent nonmagnetic $J_{\rm eff} = 0$ state as has been anticipated in pentavalent iridates (Ir-$d^4$)\cite{PhysRevLett.111.197201,PhysRevLett.112.056402}. On contrary, the electronic structure and presence of the large spin moment indicate that Hund coupling ($J_H$) dominates over SOC interactions and effective SOC could be treated as a perturbation in $S$ = 1 state. This also nicely captures the insulating ground state as schematically explained in \ref{GGA+U+SOC-PDOS}(e). Since $\frac{\mu_L}{\mu_S}$ is larger than one for Ba$_3$InRuIrO$_9$, the strong SOC limit is applicable for this system. The observed SOC-induced insulating ground state (see Fig.~\ref{GGA+U+SOC-PDOS}(c)) could be understood in terms of a $J_{\rm eff}$ = $\frac{1}{2}$ Mott-Hubbard state, as in other tetravalent iridates (Ir-$d^5$) \cite{JHalf_Kim,Na2irO3,Ba2IrO4}. As schematically explained in Fig.~\ref{GGA+U+SOC-PDOS}(f), the large octahedral crystal-field separates out the $t_{2g}$ states which has symmetry of $p$-orbitals ($l_{\rm eff}$ = 1). The strong SOC couples $l_{\rm eff}$ = 1 with $S = \frac{1}{2}$ to generate the  $j_{\rm eff} = \frac{3}{2}$ quartet and $j_{\rm eff} = \frac{1}{2}$  doublet states.  The electron correlation effects open a gap in half-filled $j_{\rm eff} = \frac{1}{2}$ as first reported in Sr$_2$IrO$_4$\cite{JHalf_Kim}. The magnitude of the gap is found to be 0.37 eV. \\
\par
\begin{figure*}[t]
    \includegraphics[width=1.99\columnwidth]{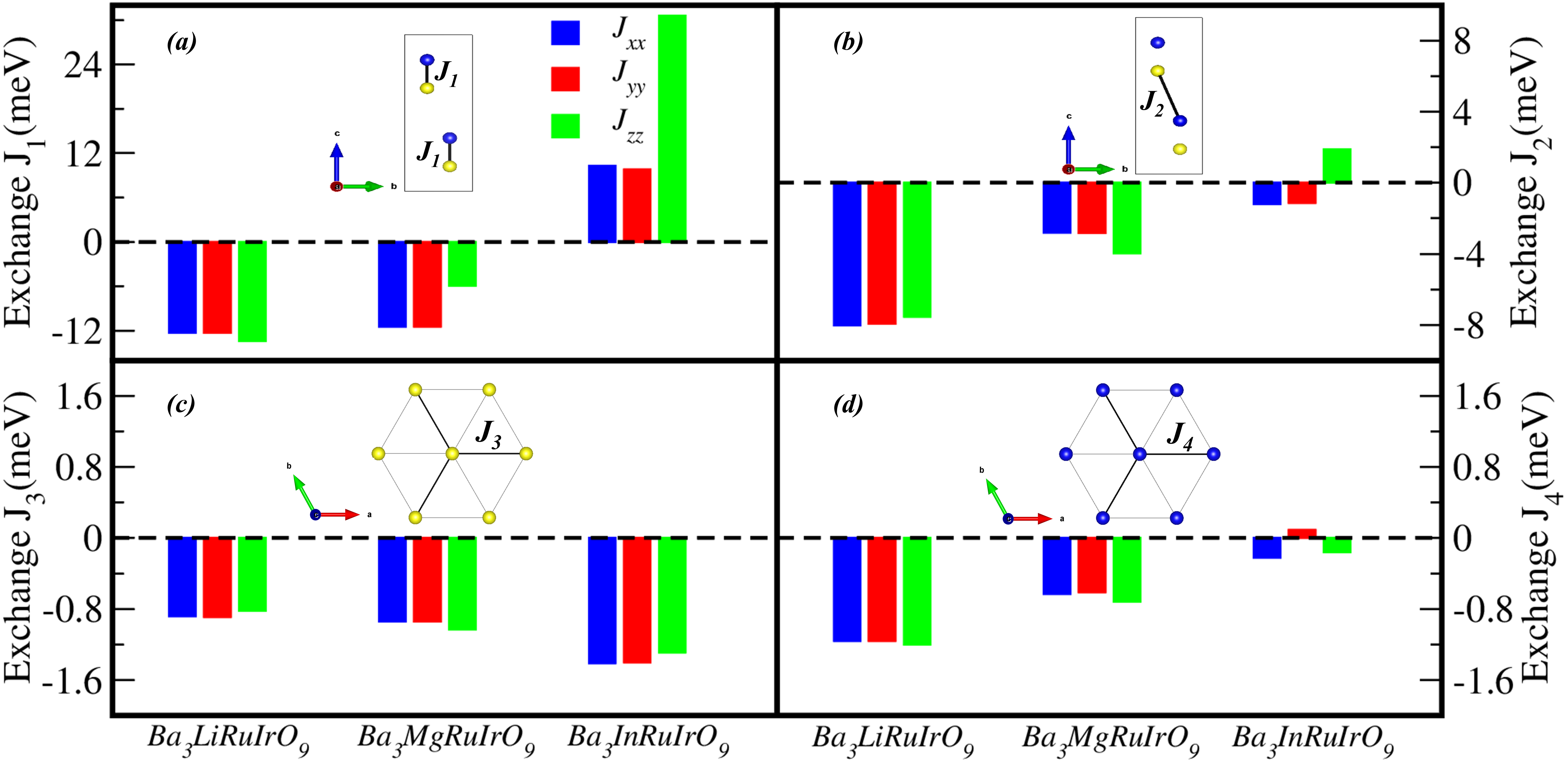}
    \caption {The diagonal elements ($J_{xx}$, $J_{yy}$, and $J_{zz}$) of the exchange interaction matrix (Eq. 2) are shown for the first four nearest neighbors (NN) (a) $1^{st}$ NN (Ir-Ru intra-dimer interaction), (b) $2^{nd}$ NN (Ir-Ru inter-dimer interaction), (c) $3^{rd}$ NN (Ru-Ru in-plane interaction), and (d) $4^{th}$ NN (Ir-Ir in-plane interaction) in Ba$_3$LiRuIrO$_9$, Ba$_3$MgRuIrO$_9$, and Ba$_3$InRuIrO$_9$. The figure reveals that the anisotropy in the diagonal components increases with the effective strength of spin-orbit coupling (SOC) in these material, as seen when moving from Ba$_3$LiRuIrO$_9$ (weak SOC) to Ba$_3$InRuIrO$_9$ (strong SOC).}
    \label{J1-schematic}
\end{figure*}
\noindent\textbf{Analyses of interatomic exchange interactions $J_{ij}$:} Next to provide a microscopic analysis of the experimental magnetic data, we computed the interatomic exchange interactions ($J_{ij}$). Following the implementation of Ref.~\cite{PhysRevB.102.115162,PhysRevB.103.174422,magnetic-force1}, we calculated J$_{ij}^{\alpha \beta}$ by mapping the computed energies from GGA+U+SOC approaches onto a  Heisenberg model as given below:
\begin{equation}
 H=-\sum_{i\neq j} e_{i}^\alpha J_{ij}^{\alpha \beta}e_{j}^{\beta}, \quad \alpha,\beta = x,y,z. 
\end{equation}
In the above equation, $J_{ij}^{\alpha \beta}$ is a second rank tensor expressed in the ${xyz}$ basis, for
which the $x$–$y$ plane is in the hexagonal $a$-$b$ plane while the $z$-axis is along the crystallographic $c$-direction and has the following form:
\begin{equation}
J_{ij}^{\alpha \beta} =
\begin{pmatrix}
J_{xx} & J_{xy} & J_{xz}\\
J_{yx} & J_{yy} & J_{yz}\\
J_{zx} & J_{zy} & J_{zz}
\end{pmatrix}
\end{equation}
The first four most dominant inter-atomic couplings, $J_1$ to $J_4$ are marked in Fig.~\ref{crystal}(b) and (c) and the estimated values are displayed in Table \ref{Isotropic} where these values are obtained from the average of diagonal components ($\frac{J_{xx}+J_{yy}+J_{zz}}{3}$). 
\par 
Our GGA+U+SOC results find that the leading term in the generalized Heisenberg Hamiltonian comes from Ir-Ru intra-dimer interaction ($J_1$) which is antiferromagnetic for Ba$_3$LiRuIrO$_9$, Ba$_3$MgRuIrO$_9$ and ferromagnetic for Ba$_3$InRuIrO$_9$. This is consistent with the total energy calculations as reported in Table~\ref{moment}. We note here that in the absence of SOC, we also computed $J_1$ and found that it is strongly antiferromagnetic for all the three systems. A dramatic change in the nature of coupling from antiferromagnetic to ferromagnetic is observed in Ba$_3$InRuIrO$_9$ only when SOC is included. The nature of these exchange interactions could be physically understood within the framework of extended Kugel-Khomskii model~\cite{kugel1982jahn}. In this model, the virtual hopping between the effective orbitals ($t_{ij}$) play the crucial role along with Coulomb interactions ($U$) and Hund's coupling ($J_H$). From the analysis of the electronic structure, we already know that Ru-$t_{2g}$ and Ir-$t_{2g}$ orbitals participate in forming magnetic moment at their respective sites and $e_g$ orbitals has no role as they appear far above Fermi energy (see Fig.~\ref{GGA+U+SOC-PDOS}) owing to large octahedral crystal-field splitting. The nominal occupations of the Ru ions are always $t_{2g}^3$ (half-filled), while the occupations of Ir are $t_{2g}^3$, $t_{2g}^4$, and $t_{2g}^5$, respectively for Ba$_3$LiRuIrO$_9$, Ba$_3$MgRuIrO$_9$, and Ba$_3$InRuIrO$_9$. The hopping being spin-conserved, up-spin electrons of Ru-$t_{2g}$ are allowed to hop to the neighboring Ir-site if the up-spin channel of the Ir-$t_{2g}$ is partially or fully empty and vice-versa. Since Ir-$t_{2g}$-states are exactly half-filled in Ba$_3$LiRuIrO$_9$ (see schematic in Fig.~\ref{GGA+U+SOC-PDOS}(d)), Ru-Ir inter-site $t_{2g}$-$t_{2g}$ virtual hoppings are possible only if they possess anti-parallel alignments, making the Ru-Ir exchange strongly antiferromagnetic. Since other spin channel of Ir is also getting occupied for Ba$_3$MgRuIrO$_9$, and Ba$_3$InRuIrO$_9$ (see schematic in Fig.~\ref{GGA+U+SOC-PDOS}(e and f)), virtual hoppings are now allowed for both parallel and anti-parallel alignments, making the net antiferromagnetic coupling relatively weaker for them. However, a ferromagnetic $J_1$ in the presence of SOC is possible if there is strong mixing between the majority and minority spin channel of the $t_{2g}$ character in the spin-orbit coupled wave-functions as discussed in Ref.~\cite{PhysRevB.99.115119}. These results again emphasise that the effective SOC is maximum in Ba$_3$InRuIrO$_9$.   
The magnitude of inter-dimer coupling between Ru and Ir ($J_2$) is strong antiferromagnetic for Ba$_3$LiRuIrO$_9$, decrease substaintially in Ba$_3$MgRuIrO$_9$ and become weak in Ba$_3$InRuIrO$_9$.  The estimated in-plane interactions within the hexagonal $a$-$b$ plane between Ru-Ru ($J_3$) come out to be antiferromagnetic. Since these interaction are antiferromagnetic in nature, they have the potential to induce magnetic frustration in these systems due to the in-plane triangular geometry as depicted in Fig.~\ref{crystal}(c). The average magnitude of $J_4$ is relatively smaller. We will discuss below how the resulting magnetic behaviours depends on these couplings. \\
\par
\noindent\textbf{Magnetic transitions using atomistic spin dynamics simulations:} To understand the experimental magnetic transitions and to validate the accuracy of computed exchange interactions, we performed atomistic spin dynamics simulations by setting up a spin Hamiltonian (see equ.~\ref{spin-hamil})  and solving it using the Landau-Lifshitz-Gilbert (LLG) \cite{spindynamics} equation as described in method section. In this spin Hamiltonian, we consider all computed magnetic couplings, including MAE. We note that the total obtained MAE (see Table~\ref{MAE}) is distributed among Ru and Ir according to the values of their computed spin-orbit coupling constants which are129 meV for Ru and 455 meV for Ir. 
The calculated temperature dependent magnetic specific heat plots are displayed in Fig~\ref{ASD_results_1}. These results demonstrate that the ordering temperature ($T_N$) of Ba$_3$LiRuIrO$_9$ and Ba$_3$MgRuIrO$_9$ comes out to be 152 K and 93 K respectively. The good agreement between the experimental ordering temperature (180 K for Ba$_3$LiRuIrO$_9$ and 100 K for Ba$_3$MgRuIrO$_9$~\cite{lufaso2005crystal}) and the corresponding theoretically estimated values demonstrate the accuracy of our computed parameters that are used in constructing the spin Hamiltonian. Thus, we are able to explain the change in transition temperature in these two systems. To interpret the underlying cause for the change in $T_C$, we looked at the effective interactions which is $z_iJ_i$, $z_i$ being the number of nearest neighbor (NN) magnetic ions. We have displayed these values in Table~\ref{Isotropic}. For Ba$_3$LiRuIrO$_9$, the effective inter-dimer coupling (3$J_2$) is found to be almost twice as large compared to effective intra-dimer coupling ($J_1$). The in-plane frustrated interactions such as 6$J_3$ and 6$J_4$ are relatively weaker. Thus, the observed long range magnetic ordering in Ba$_3$LiRuIrO$_9$ could be attributed to the strong Ir-Ru intra and inter-dimer couplings. For Ba$_3$MgRuIrO$_9$, $J_1$ comes out to be the most dominating coupling and effective $J_2$ has similar magnitude. In-plane effective interactions arising due to $J_3$ and $J_4$ being anti-ferromagnetic in nature try to hinder the magnetic ordering but are comparatively weak than $J_1$ and $J_2$. Our results find an overall reduction of 40$\%$ in the effective magnetic coupling ($J_1$+3$J_2$+6$J_3$+6$J_4$) compared to Ba$_3$LiRuIrO$_9$, indicating an equivalent reduction in transition temperature according to the mean field theory. Interestingly,  experimental results~\cite{lufaso2005crystal} and our spin-dynamics simulations also find a similar reduction in transition temperatures. In both these systems the dominating magnetic coupling terms are intra and inter-dimer Ru-Ir exchange which are responsible for the long range orderings. However, an interesting change is observed in Ba$_3$InRuIrO$_9$ where the second dominating effective coupling arise from in-plane Ru-Ru exchange interactions (6$J_3$) which is just half of the ferromagnetic intra-dimer coupling ($J_1$) and almost three times larger than the effective inter-dimer couplings. We note here that $J_3$ is geometrically frustrated in nature and this could play a dominant role in the ground state magnetism of this material. In the later part of the manuscript, we will analyze in detail the reason for the absence of long range ordering in Ba$_3$InRuIrO$_9$~\cite{lufaso2005crystal} and the possible emergence of novel magnetic ground state in this system. \\
\begin{table}
\centering
\caption{The computed Dzyaloshinskii-Moriya interactions ($D_{ij}$) and symmetric anisotopic exchange ($C_{ij}$) corresponding to different NN in Ba$_3$InRuIrO$_9$.}
\label{DM}
\begin{tabular}{|cc|cc|}
\hline
\multicolumn{2}{|c|}{}                     & \multicolumn{2}{c|}{$Ba_{3}InRuIrO_{9}$}                                  \\ \hline
\multicolumn{1}{|c|}{Neighbor} & Bonds     & \multicolumn{1}{c|}{$D_{ij}(meV)$}              & $C_{ij}(meV)$             \\ \hline
\multicolumn{1}{|c|}{$1^{st}$}      & Ir-Ru     & \multicolumn{1}{c|}{(0.84, 0.21, 0.10)}    & (1.19, 0.01, 0.45)   \\ \hline
\multicolumn{1}{|c|}{$2^{nd}$}      & Ir-Ru     & \multicolumn{1}{c|}{(-0.38, 0.59, -0.04)}  & (-0.25, -0.14, 0.65)  \\ \hline
\multicolumn{1}{|c|}{$3^{rd}$}      & Ru-Ru     & \multicolumn{1}{c|}{(-0.01, -0.06, 0.04)}  & (0.02, 0.00, 0.00)  \\ \hline
\multicolumn{1}{|c|}{\multirow{1}{*}{$4^{th}$}} & Ir-Ir & \multicolumn{1}{c|}{(-0.02, 0.09, 0.10)} & (0.23, 0.00, 0.04) \\ \hline
\end{tabular}
\end{table}
\par 
\noindent\textbf{The SOC as a source of exchange anisotropy:}
Before analyzing the ground-state magnetism of the most fascinating materials in this series, namely Ba$_3$InRuIrO$_9$, we discuss another interesting aspect of exchange interactions that is found when we looked at the diagonal components of $J_{ij}$ tensor. We have schematically shown the three components of $J_1$ in Fig. \ref{J1-schematic}(a). Our results suggest that the strength of the interaction along the $z$-axis ($J_1^{zz}$) is different compared to the $x$-$y$-plane ($J_1^{xx}$ = $J_1^{yy}$), indicating an anisotropic $xxz$ type Heisenberg model for these systems as far as the most-dominant intra-dimer interaction $J_1$ is concerned. The anisotropy parameter could be defined as $\Delta = \frac{J_{zz}-J_{xx}}{J_{xx}}$. The value of $\Delta$ is found to be 0.09 for Ba$_3$LiRuIrO$_9$, 0.49 Ba$_3$MgRuIrO$_9$ and 1.97 for Ba$_3$InRuIrO$_9$. We see that the $\Delta$ is negligible for Ba$_3$LiRuIrO$_9$ where effective SOC is very poor and gradually increases for the other two systems. Hence, we see an interesting phenomena that spin-orbit coupling is not only responsible for the finite off-diagonal components of $J_{ij}$ tensor but also induce anisotropy in its diagonal components. A similar anisotropy in diagonal components is also observed in $J_2$ as schematically presented in Fig. \ref{J1-schematic}(b). Here, the value of $\Delta$ varies as 0.05 for Ba$_3$LiRuIrO$_9$, 0.41 Ba$_3$MgRuIrO$_9$ and 2.56 for Ba$_3$InRuIrO$_9$. This further strengthen our argument that the strong SOC effect is observed in Ba$_3$InRuIrO$_9$ where Ir ions exhibit 4+ charge state. In Fig. \ref{J1-schematic}(c), diagonal components for Ru-Ru inplane interactions are shown. Here, $\Delta$ value is found to be very negligible i.e., 0.07, 0.09, 0.09 for Li, Mg and In system respectively, suggesting weak anisotropy for Ru-Ru in-plane hexagonal network. This behaviour is seen because poor SOC effects are expected for Ru ions due to the nominal occupancy of $t_{2g}^3$ throughout. Fig. \ref{J1-schematic}(d) shows diagonal componets for Ir-Ir in plane interactions. Here, $\Delta$ value is found to be 0.03 for Ba$_3$LiRuIrO$_9$, 0.14 Ba$_3$MgRuIrO$_9$ and 0.27 for Ba$_3$InRuIrO$_9$. This anisotropy in the diagonal components of $J_{ij}$ further confirms that SOC effect enhances from poor to moderate to strong as nominal occupation of Ir changes from $t_{2g}^3$ to $t_{2g}^4$ to $t_{2g}^5$. In other words, these results also demonstrate that spin-orbit coupling not only induces off-diagonal components in $J_{ij}$ tensor, but also responsible for the anisotropy in the diagonal components. \\
\par
\noindent\textbf{Magnetic ground state of Ba$_3$InRuIrO$_9$:}
We have now established that the nominal charge state of Ir and consequently the occupation of $t_{2g}$ states play a very crucial role in the electronic structure and magnetism of Ir-based compounds. Our detailed analyses provides compelling evidences that magnetic frustration and effective SOC are strongest in Ba$_3$InRuIrO$_9$ where Ir is in $d^5$ state like Sr$_2$IrO$_4$ ~\cite{Sr2IrO4-PRL}, Na$_2$IrO$_3$~\cite{Na2IrO3-yamaji2014first} etc.  It is well known that magnetic frustration and SOC combined can give interesting consequences, including the emergence of exotic ground states, such as QSL or spin glasses as reported for many Ru and Ir based materials (i.e. Na$_2$IrO$_3$, Li$_2$IrO$_3$, NaRuO$_2$, RuCl$_3$~\cite{takagi2019concept,kim2016revealing,RuCl3} etc). In addition, Ba$_3$InRuIrO$_9$ also possess other important characteristics such as insulating ground state, a paramagnetic magnetic susceptibility with negative Curie temperature~\cite{lufaso2005crystal}. These signatures point toward the possibility of a QSL ground state, making it necessary to analyse the relevant magnetic model for this  system. Such magnetic ground states are of interest both from a fundamental perspective to advance our understanding of quantum magnetism, and for potential applications in fields such as spintronics and quantum computing. Hence the rest of the manuscript has been focused on deriving a spin model for Ba$_3$InRuIrO$_9$. 
\begin{figure*}
    \includegraphics[width=1.99\columnwidth]{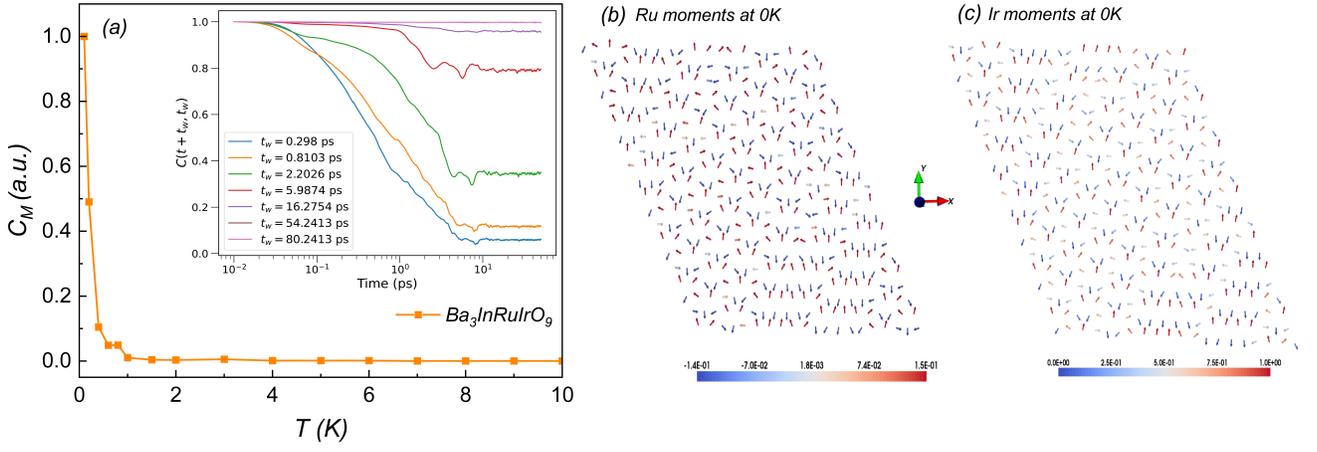}
    \caption{(a) Finite-temperature heat capacity calculated from atomistic spin dynamics simulations for Ba$_3$InRuIrO$_9$ , revealing the absence of a sharp peak that would indicate a long-range magnetic transition. The inset shows the spin autocorrelation function $C(t+t_w,t_w)$ for different waiting times ($t_w$). The decay of the autocorrelation function to a stable value within approximately 100 ps suggests complex magnetism but rules out conventional spin-glass behavior. Panels (b) and (c) visualize one of the many possible zero-temperature spin configurations for the Ru and Ir layers, respectively. The absence of a clear pattern in the spin orientations further confirms the lack of long-range magnetic order.}
   \label{ASD_results-2}
\end{figure*}
\par 
In general, the spin Hamiltonian, $H_{spin}$ consists of a sum of one-spin and two-spin terms as expressed below: 
\begin{equation}
\label{spin-hamil}
H_{spin} = -\sum_{i\neq j}\left(J_{ij}\overrightarrow{S_i}\cdot\overrightarrow{S_j} + \overrightarrow{S_i}C_{ij}\overrightarrow{S_j}+ D_{ij}\cdot[\overrightarrow{S_i}\times\overrightarrow{S_j}] \right) - \sum_{i} \epsilon_{An}^{i} ({S_{i}^{z}})^{2}
\end{equation}
Here the first term represent Heisenberg exchange which is bilinear in spins, the second term is the symmetric anisotropic exchange while third term represents the antisymmetric anisotropic exchange also known as Dzyaloshinskii-Moriya interactions (DMI). The only one-spin term  is single-ion anisotropy, the last term of the Hamiltonian. All the two-spin terms are related to the magnetic exchange tensor, J$_{ij}^{\alpha \beta}$. We have already discussed about the diagonal terms.  Further we looked into the off-diagonal components of Ir-Ru and Ir-Ir magnetic couplings in order to calculate the Dzyaloshinski–Moriya (DM) interactions ($D_{ij}$) and symmetric anisotropic exchange ($C_{ij}$). 
\par 
The components of $D_{ij}$ and $C_{ij}$ are computed from the off-diagonal elements of the $J_{ij}$ matrix by using the following equations 
\begin{equation}
   D_{ij}^{x}= \frac{1}{2}\{J_{ij}^{yz}-J_{ij}^{zy}\},\quad D_{ij}^{y}= \frac{1}{2}\{J_{ij}^{xz}-J_{ij}^{zx}\},\quad D_{ij}^{z}= \frac{1}{2}\{J_{ij}^{xy}-J_{ij}^{yx}\},
\end{equation}
\begin{equation}
   C_{ij}^{x}= \frac{1}{2}\{J_{ij}^{yz}+J_{ij}^{zy}\},\quad C_{ij}^{y}= \frac{1}{2}\{J_{ij}^{xz}+J_{ij}^{zx}\},\quad C_{ij}^{z}= \frac{1}{2}\{J_{ij}^{xy}+J_{ij}^{yx}\}.
\end{equation}
As displayed in Table \ref{DM}, the $D_{ij}$ and $C_{ij}$ for the 1st and 2nd NN (Ru-Ir intra and inter-dimer) are strong. 
Using the information on the magnetic interactions (exchange and DMI) and MAE discussed above, we performed atomistic spin dynamics simulations for Ba$_3$InRuIrO$_9$ using LLG equation as discussed before. Our results of magnetic specific heat as displayed in Fig.~\ref{ASD_results-2}(a) do not find any signatures of long-range magnetic order which is in good agreement with the experimental data\cite{lufaso2005crystal}. This is markedly different from the cases of Ba$_3$LiRuIrO$_9$ and Ba$_3$MgRuIrO$_9$ where sharp peaks in the heat capacity was observed (see Fig.~\ref{ASD_results_1}). 
In order to investigate the relaxation properties of the system we have sampled the time-displaced spin-spin autocorrelation function
\begin{equation}
   \label{eq:ac}
   C_0(t+t_w,t_w) = \langle \overrightarrow{S_i}(t+t_w)\cdot\overrightarrow{S_i}(t_w) \rangle
\end{equation}
in simulations using a temperature quenching protocol. We note first that for a system that has fully relaxed in thermal equilibrium, the autocorrelation function is time translationally invariant, i.e. for all waiting times $t_w$ the autocorrelation function $C_0(t+t_w,t_w) = C_0(t)$. For a slowly relaxing system such as a system in spin glass phase, the relaxation can proceed over many orders of magnitude of time, a regime which is referred to as an aging regime \cite{spindynamics}.
\par
To explore the possibility of a spin glass phase in Ba$_3$InRuIrO$_9$ we have used a two-stage quenching protocol. At time $t<0$ we have equilibrated the system at the temperature $T=10$ K, which according to Fig.~\ref{ASD_results-2}(a) corresponds to a temperature where the spins are in a disordered phase. At time $t=0$ the temperature is abruptly quenched to $T=0.1$ K and the system is evolved in spin dynamics simulations using a Gilbert damping $\alpha=0.01$. Sampling the autocorrelation for a set of waiting times logarithmically distributed in time, we obtained a set of data shown in the inset of Fig.~\ref{ASD_results-2}(a). For each waiting time, the autocorrelation starts of at unity value. For the lowest waiting times $t_w=0.298$ ps the autocorrelation decay saturates within 10 ps to a value around 0.06, and for $t_w=0.8103$ ps the autocorrelation function is stabilizing around 0.12. Also for the intermediate waiting times $t_w=2.2026$, 5.9874, and 16.2754 ps the decay to a stable value happen over around 10 ps, with the autocorrelation for $t_w=16.2754$ ps settling at 0.96. At the second largest waiting time $t_w=54.2413$ ps the system has relaxed to a state where the spins precess with small cone angle in their local magnetic field, with the autocorrelation oscillating at values around 0.997. The autocorrelation for the highest waiting time $t_w=80.2413$ ps is on top of the $t_w=54.2413$ ps autocorrelation, indicating that the spin system has fully relaxed. Full relaxation happening on the time scale of 100 ps is within a regime of regular relaxation for a spin system, and not in the regime of aging dynamics of a spin glass for which aging can proceed over multitudes of times scales. The data in Fig.~\ref{ASD_results-2}(a) hence point to complex magnetism as T approaches 0 K, and that conventional spin-glass behavior can be ruled out.
\par
The orientation of Ru and Ir spin moments are also displayed in Fig.~\ref{ASD_results-2}(c). The absence of any long range ordering is clearly visible from these plots. To understand the microscopic origin of the absence of long-range ordering, we investigated all the possible sources of magnetic frustration in this material. The strong magnitudes of $D_{ij}$ and $C_{ij}$ imply that the off-diagonal components of the $J_{ij}$ matrix are significant and play an important role in determining the magnetism of these materials. The $D_{ij}$ and $C_{ij}$ for 1st and 2nd NN are very strong. As expected, the magnitudes of  these quantities for the 3rd NN interactions (Ru-Ru bonds in $a-b$-plane) are very small due to the quenching of SOC in Ru-ions. The value of $C_{ij}$ for the 4th NN (Ir-Ir in-plane frustrated couplings) also comes out to quite significant. This is due to the presence of strong off-diagonal terms and some of them are larger than the diagonal components. For instance the off-diagonal elements of $J_4$ are $J_{xy}$= 0.14 meV, $J_{xz}$=0.09 meV, $J_{yz}$= 0.21 meV, $J_{yx}$= -0.06 meV, $J_{zx}$= -0.09 meV and $J_{zy}$= 0.25 meV. Such strong off-diagonal components have potential to induce spin-liquid ground state as discussed in Ref.\cite{PhysRevLett.118.147204,PhysRevLett.102.017205}. 
\par
The realization of strong anisotropic interactions as a consequence of the interplay between spin–orbit coupling (SOC), crystal field splitting and Hund’s coupling\cite{PhysRevLett.102.017205} further motivate us to investigate if Ba$_3$InRuIrO$_9$ also belongs to a novel class of Kitaev materials\cite{KITAEV20062,Takagi2019,qsl-nature}. 
In analogy to the intensively studied $t_{2g}^5$ triangular-lattice material NaRuO$_2$\cite{NaRuO2-QSL,NaRuO2-QSL1} that is found to host a quantum spin liquid ground state, the strong SOC of iridium, combined with the geometry of face-sharing ligand octahedra, is expected to form a $J_{\rm eff} = \frac{1}{2}$ magnet with finite Kitaev interaction\cite{PhysRevLett.102.017205}. Hence  we adopted the method of Ref\cite{CrI3-QSL} to compute the Kitaev term. We first diagonalized the $J_4$ matrix to obtain its eigenvalues which are denoted $J_4^X$, $J_4^Y$, $J_4^Z$. We found that $J_4^X$= 0.19 meV,  $J_4^Y$= -0.19 and $J_4^Z$= -0.30 meV. Now, all the two spin-terms of $H_{\rm spin}$ for $J_4$ in this orthogonal $\{XYZ\}$ basis could be rewritten in a more simplified form as follows:
\begin{equation}
H_{ex} = \sum_{ij}\left(J\overrightarrow{S_i}\cdot\overrightarrow{S_j} + KS_i^{Z}S_j^{Z}\right)
\end{equation}
where $J = J_X$ is the isotropic exchange coupling and $K = J_Z - J_X$ is the so-called Kitaev interaction that characterizes the anisotropic contribution. The Kitaev term comes out to be K=-0.5 meV. A similar analysis for the non-frustrated Ru-Ir intra-and inter-dimer couplings (J$_1$, J$_2$ ) also give rise to a even stronger anisotropic magnetic behavior as schematically presented in Fig.~\ref{J1-schematic}. In addition, we also emphasize that Ba$_3$InRuIrO$_9$ exhibit a strong easy plane magnetocrystalline anisotropy which could induce exchange frustration\cite{TREBST20221} due to the fact that bond-directional interactions with competing quantization axes cannot be satisfied simultaneously. 
Thus, our results provide evidences that this system is a possible candidate for QSL~\cite{ANDERSON1973153}.
\section{CONCLUSION}
In conclusion, we theoretically studied the electronic structure and magnetism of mixed Iridium-Ruthenium triple Perovskites with the chemical formula Ba$_3$MRuIrO$_9$ (M = Li, Mg and In). Our results suggest that the modulation of iridium charge states emerges as a key factor in tuning the strength of spin-orbit coupling and the anisotropy of exchange interactions. In Ba$_3$LiRuIrO$_9$, we found that both Ru and Ir have half filled $t_{2g}$ states (nominal $d^3$ occupation) resulting in the quenching of orbital moments and a poor strength of effective SOC. The electronic structure of this system can be described in terms of a band insulator. For Ba$_3$MgRuIrO$_9$, Ru is again found in nominal $d^3$ occupation while Ir exhibits 5+ charged state acquiring a nominal $d^4$ configuration. As a result, finite orbital moment is observed at Ir site which manifest a moderate SOC strength that can be treated perturbatively in $S$ = 1 state. The prediction of $J = 0$ state in pentavalent iridates does not appear to be correct due to the moderate strength of SOC. In case of Ba$_3$InRuIrO$_9$, Ir occupancy is further changed to nominal $d^5$ configuration while Ru maintains its nominal $d^3$ occupation. As a result, the largest orbital moment is observed at the Ir site resulting in large SOC effects. Hence, Ba$_3$InRuIrO$_9$ is found to be a candidate of relativistic $J_{\rm eff} = \frac{1}{2}$ Mott-Hubbard insulator. Thus our results are able to explain the role of Ir-charge state in tuning the strength of effective SOC and provide a compelling evidence for the stability limit of $J_{\rm eff}$ state. We compute the diagonal and off-diagonal components of inter-atomic exchange tensor with an aim to describe experimental magnetic data and also to provide spin-model for this interesting class of materials. Our spin dynamics study, solving the Landau-Lifshitz-Gilbert equation of motion, explain the long-range ordering and reproduce the measured value of transition temperature of Ba$_3$LiRuIrO$_9$ and Ba$_3$MgRuIrO$_9$. These calculations also explain the absence of long range ordering in Ba$_3$InRuIrO$_9$ and that a complex magnetic phase transition sets in as temperature approaches 0 K. Further we analyzed the diagonal and off diagonal components of $J_{ij}$ and mapped it to a Heisenberg-Kitaev model. Our analyses provide evidences of strong magnetic frustration in Ba$_3$InRuIrO$_9$, that does not favor a spin-glass state, and showed that this is a potential candidate for quantum spin liquid. We reported the presence of finite bond-directional interactions. Thus, our study establishes a theoretical framework where the interplay of strong spin-orbit coupling, crystal field effects, and bond-directional exchange interactions can possibly stabilize a quantum spin liquid (QSL) ground state in iridium-based oxides. A comparison of iso-structural iridates with different Ir-charge states nicely demonstrate the electronic conditions for this interplay to drive the deviations from conventional magnetic ordering, facilitating the emergence of exotic quantum phases. This important theoretical understanding not only deepens insights into the conditions required for quantum phase transitions but also provides a road-map for identifying other 5$d$ transition metal oxides with similar attributes. Future experimental studies can leverage these insights to design and engineer materials with tailored quantum phases. 
\par
\section{ACKNOWLEDGEMENTS}
S.K.P acknowledges support from DST SERB grant (CRG/2023/003063). O.E. acknowledges financial support from the Swedish Research Council (VR), the European Research Council (ERC grant 854843-FASTCORR), eSSENCE, STandUPP, WISE - Wallenberg Initiative Materials Science funded by the Knut and Alice Wallenberg (KAW) Foundation and KAW projects (KAW 2022.0108 and KAW 2022.0252). The computations were enabled by resources provided by the National Academic Infrastructure for Supercomputing in Sweden (NAISS), partially funded by the Swedish Research Council through grant agreement no. 2022-06725.

\end{document}